\documentclass[10pt,a4]{report} 
\usepackage{subfigure}  
\usepackage{fancyhdr}  

\usepackage{amsmath}
\usepackage{amssymb}

\usepackage{makeidx}

\newif\ifpdf
\ifx\pdfoutput\undefined
   \pdffalse
\else
   \pdfoutput=1
   \pdftrue
\fi

\ifpdf
  \usepackage[
  pdftex,
  hyperindex,
  pagebackref,
  colorlinks,
  linkcolor=blue,
  menucolor=blue,
  pagecolor=blue,
  urlcolor=blue,
  citecolor=blue,
  pdftitle={},
  pdfauthor={Marco Caparrini},
  pdfsubject={},
  pdfcreator={LaTeX},
  pdfkeywords={},
  ]{hyperref}
  \usepackage{url}
  \usepackage[pdftex]{graphicx}
  \usepackage[pdftex]{color} 

\else
\usepackage{graphicx}  

\usepackage[dvips]{color} 

\fi
\usepackage{xfig}
\graphicspath{{figures/}}

\usepackage[latin1]{inputenc}    

\catcode`\@=11

\catcode`\@=11
\def\section{\@startsection{section}{1}{\z@}{3.5ex plus 1ex minus
   .2ex}{2.3ex plus .2ex}{\large\bf}}

\def\eqnarray{\let\@currentlabel=\theequation\refstepcounter{equation}
    \global\@eqnswtrue
    \global\@eqcnt\z@\tabskip\@centering\let\\=\@eqncr
    $$\halign to \displaywidth\bgroup\@eqnsel\hskip\@centering
      $\displaystyle\tabskip\z@{##}$&\global\@eqcnt\@ne
       \hfil${{}##{}}$\hfil
      &\global\@eqcnt\tw@ $\displaystyle\tabskip\z@{##}$\hfil
       \tabskip\@centering&\llap{##}\tabskip\z@\cr}
\def\lefteqn#1{\hbox to 4\arraycolsep{$\displaystyle #1$\hss}}

\def\thesection{\arabic{section}.}

\def\appendix{\setcounter{section}{0}
        \def\thesection{Appendix.}
        \def\theequation{\Alph{section}.\arabic{equation}}}

\long\def\@makefntext#1{\parindent 0cm\noindent
\hbox to 1em{\hss$^{\@thefnmark}$}#1}

\newcommand\unmarkfootnote[1]{%
  \begingroup 
    \let\@makefntext\noindent
    \footnotetext{#1}%
  \endgroup}

\newcommand{\captionfonts}{\footnotesize}

\makeatletter  
\long\def\@makecaption#1#2{%
  \vskip\abovecaptionskip
  \sbox\@tempboxa{{\captionfonts #1: #2}}%
  \ifdim \wd\@tempboxa >\hsize
    {\captionfonts #1: #2\par}
  \else
    \hbox to\hsize{\hfil\box\@tempboxa\hfil}%
  \fi
  \vskip\belowcaptionskip}
\makeatother   

\newsavebox{\fmbox}
\newenvironment{fmpage}[1]
{\begin{lrbox}{\fmbox}\begin{minipage}{#1}}
    {\end{minipage}\end{lrbox}\fbox{\usebox{\fmbox}}}

\def\str{STARLAB Barcelona SL\space}

\begin{document}
\pagestyle{empty}
                                    
\thispagestyle{empty}
\vspace{.5in}

\begin{center}

{\Huge\bf PARIS Alpha}\\\vspace{.15in}
{\Large\ ESA/ESTEC Contract No. 14285/85/nl/pb     }\\\vspace{.15in}
{\Huge\bf PARIS  altimetry \\\Huge\bf with L1 frequency data  \\[0.4cm]\Huge\bf from the Bridge 2 campaign}\\\vspace{.2in}
{\Large\bf CCN3-WP3 Technical Report }\\\vspace{.4in}
{\large\bf Abridged version }\\\vspace{.4in}
\unmarkfootnote{\copyright \str}

\vspace{.9cm}

\includegraphics[width=8cm]{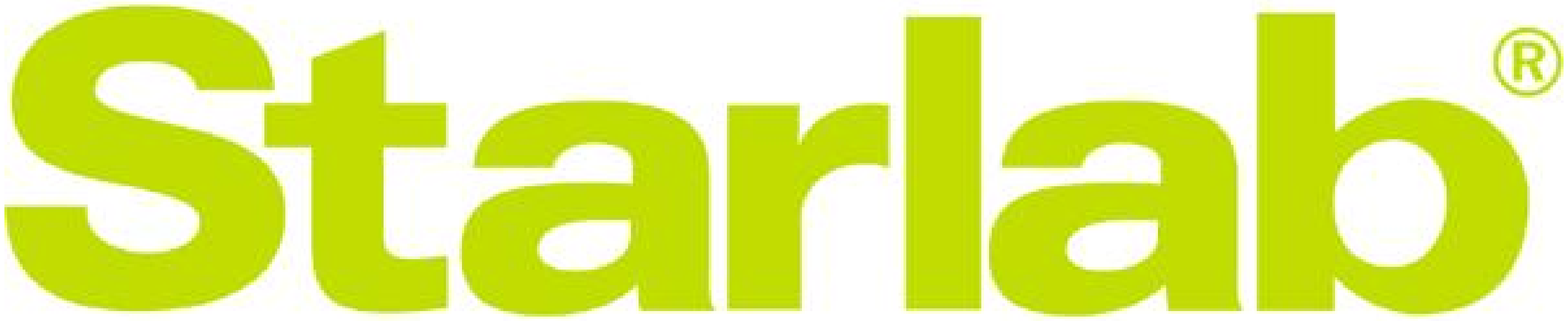} \\ 
\vspace{.9cm}

{\large\bf ESA/ESTEC Technical Manager: M. Martin-Neira, TOS-ETP   }\\

\vspace{2cm}
   {Prepared by \\
G. Ruffini, M. Caparrini, L. Ruffini \\
 
    Approved by Giulio ~R{uffini} ({\it giulio@starlab-bcn.com})\\
    {\small\it Starlab Barcelona, April 26th, 2002}\\
        {\small\it Edifici de l'Observatori Fabra, C. de l'Observatori s.n.}\\
{ \small\it Muntanya del Tibidabo, 08035 Barcelona, Spain \\ }
        }
\end{center}

\vfill{ \small
\begin{fmpage}{\textwidth}
\begin{center}
EUROPEAN SPACE AGENCY\\
CONTRACT REPORT\\
The work described in this report was done under ESA contract.\\
Responsibility for the contents resides in the author or organisation that prepared it.
\end{center}
 \end{fmpage}
}

\begin{abstract}
A portion of 20 minutes of the GPS signals collected during the Bridge 2 experimental campaign, performed by ESA, have been processed.\medskip

 An innovative algorithm called \sc Parfait \rm,  developed by Starlab and implemented within Starlab's GNSS-R Software package STARLIGHT (STARLab Interferometric Gnss Toolkit), has been successfully used with this set of data.\medskip

A comparison with tide values independently collected and with differential GPS processed data has been performed. We report a successful PARIS phase altimetric measure of the Zeeland Brug over the sea surface with a rapidly changing tide, with a precision better than 2 cm.
\end{abstract}



\section{The GPS reflected field retrieved by GPS processing}
\label{ch:field_with_GPS}
\subsection{Introduction}
As reported in \cite{ruffini2001a}, during the last decade many experimental campaigns to recover GPS reflected data have been successfully organised (a partial list in Table (\ref{tab:GNSSR_experiments})).

The data which have been used in the work described in this paper were collected during the \it Bridge-2 experiment \rm \cite{belmonte2002}. \\

This article covers the following issue:
\begin{itemize}
\item Analysis of the direct and reflected signals and EM field extraction.
\item Analysis of the reflected field behaviour.
\item PARIS phase altimetry.
\item Analysis of the altimetric results.
\end{itemize}

\begin{table}
\hspace{-1cm}
 \label{tab:GNSSR_experiments}
\centering{

   \begin{tabular}[tb]{|p{1.6cm}c|p{1cm}|c|p{7cm}|}
     \hline
 Main \hspace{5mm} Author & & Affiliation & Date & -Note \\
        \hline \hline
  Garrison &\cite{garrison1996} & JPL   & 1996  & A normal receiver was used. Demonstrated reception/tracking  of reflected signals.  Concluded  that more complex receiver is needed.  \\
       \hline          
   ${\mbox{Martin-Neira}}$ Caparrini&\cite{martin-neira2001}\cite{caparrini1998}& ESA & 1997  & PARIS altimetric concept applied GNSS-R collected from a bridge.  C/A code only could be used for correlation leading to an altimetric accuracy in the order of 3 meters (1\% of the chip length).   \\
       \hline          
  Garrison & \cite{garrison1998}& JPL & 1997 &  Widening of correlation function in rough seas demonstrated. \\
       \hline          
  Komjathy &  \cite{komjathy1998}& & 1998 &  Aircraft experiments, 3-5 km altitude. \\
       \hline          
 LaBrecque & \cite{labrecque1998}& & 1998 &  The first spaceborne observation of GPS signals reflected from the ocean surface.\\
       \hline          
 Cardellach Ruffini Garrison& \cite{cardellach2001}\cite{garrison2000}& IEEC & 1999 & Balloon experiment. Successful detection of reflected signals at 38 km of height.   \\
        \hline
 Cardellach Ruffini& \cite{cardellach2001a}& IEEC & 1999 & Aircraft experiment.   \\
       \hline
 Ruffini \hspace{5mm}Caparrini & \cite{caparrini2001} & & 2000 & The set of GPS-R data collected from a drilling platform has been analysed.\\
     \hline
 Armatys & \cite{armatys2000} & & 2000 & Wind speed and directions obtained from reflected GPS signals are compared to the SeaWinds scatterometer on-board QuikSCAT.\\
     \hline
 Garrison & \cite{garrison2000a} & JPL & 2000 & With GPS-R airborne data, retrieval of the wind speed with a bias of less than 0.1 m/s and with a standard deviation of 1.3 m/s.\\
     \hline
Zavorotny & \cite{zavorotny2000}&  & 2000 & Comparison of experimental and theoretical waveforms.\\
     \hline
Zuffada & \cite{zuffada2000}& JPL & 2000 & Lakeside experiment, with an almost local planar surface (no roughness).\\
     \hline
${\mbox{Martin-Neira}}$ & \cite{pipaer2000}\cite{pipaer2000a}& ESA& 2000 & The experiment was designed to test some key issues in the PARIS Interferometric Processor (PIP) concept. The PIP concept is based on the use of dual-frequency carrier measurements to exploit the correlations in the scattered signals at similar frequencies.\\
     \hline
${\mbox{Martin-Neira}}$ & \cite{belmonte2002}& ESA & 2001 & The experimental campaign which is the object of this work.\\
     \hline
Cardellach Germain Caparrini&\cite{cardellach2002} \cite{cardellach2002b}\cite{cardellach2002c}\cite{germain2002}& IEEC STARLAB& 2001 & GPS-R data collection from airborne platform. Campaign performed within the  ESA/ESTEC project ``PARIS Alpha''. Data processed under ESA/ESTEC projects ``PARIS Alpha'' and ``OPPSCAT 2''.\\
      \hline
STARLAB & & STAR-LAB & 2002 & GPS-R data collection from airborne platform. Data processing currently performed within the  ESA/ESTEC project ``PARIS Gamma''.\\
      \hline
\end{tabular}}
\vspace{.5cm}
\caption{Representative GNSS-R experiments (1996-2002).}
\end{table}

\subsection{Phase retrieval}
The importance of retrieving the phase of the EM field backscattered by the sea surface, in an Earth observation application, is related to the possibility of accurately estimate the delay of the reflected signal w.r.t the direct one, i.e. for altimetric purposes as well as (\cite{ruffinipipaer2000}) is related to the sea surface conditions.\\
Clearly, in order to collect the phase of the EM field, the complex signal has to be considered. In the current experiment, only the in-phase component of the signal is sampled and stored. It is necessary than to generate locally a quadrature component. This is done as presented in Figure (\ref{fig:downconvertion}).
Let us represent the signal received at the antenna input \footnote{This represent only the C/A code part. The P code part can be neglected thanks to the subsequent correlation of the signal with replicas of the C/A code. The two code are in fact orthogonal.} as
\begin{equation}
  s(t)=C(t)D(t)cos((\omega_{L_1}+\omega_d)t)
\end{equation}
where $C(t)$ represent the C/A code, $D(t)$ the navigation code, $\omega_{L_1}$ the L1 carrier frequency, and $\omega_d$ the actual Doppler frequency. After modulation with a local oscillator of frequency $\omega_{L_1}-\omega_{IF}$ and low-pass filtering, the signal will have a residual carrier at $\omega_d+\omega_{IF}$. This signal is then mixed with to phasors at frequency $\omega_{IF}+\tilde{\omega}_d$, where $\tilde{\omega}_d$ is the available estimation of the Doppler frequency for the satellite under investigation. These two phasors are relatively delayed by $\pi/2$ radians, i.e. they are in quadrature.
The result of such a mixing is a signal with two frequency component: $\Delta\omega_d$, the error in the Doppler frequency estimation, and $2(\omega_{IF}+\omega_d)-\Delta\omega_d$, a spurious frequency which is not possible to filter out and that is considered as a noise component.

With the assumption that the navigation bit $D_k$ is constant during the integration time (always true if $T_{i}=1$ ms, in the other cases it is true if the code is aligned with the navigation bit and the coherent integration time is a sub-multiple of 20 ms) and that the component of the signal at higher frequency does not correlate with the clean replica of the C/A code (thanks to the PRN code properties), the \em i-th \rm sample of the correlation between the in-phase component of the signal and the clear replic can be written as
\begin{equation}
I_{i}\,=\,\frac{1}{2} D_k R_i\sum_{(i-1)N_E+1}^{i N_E}\, \cos\left(\Delta\omega_{d_k} t_k \right)
\label{eq:inphase_corr2}
\end{equation}
and for the in-quadrature component:
\begin{equation}
Q_{i}\,=\,\frac{1}{2} D_k R_i\sum_{(i-1)N_E+1}^{i N_E}\, \sin\left(\Delta\omega_{d_k} t_k \right)
\label{eq:inquad_corr2}
\end{equation}
Considering that during an integration time interval the value of $\Delta\omega_{d_k}$ is constant, joining the two signal components and performing the summation, the complex  \em i-th \rm sample of the correlation writes
\begin{equation}
C_{i}\,=\,\frac{1}{2} D_k R_i \,e^{-j\Delta\omega_{d_i}T_E\left(i-1\right)}e^{-j\Delta\omega_{d_i}\frac{T_s}{2}} \frac{\sin\left(\frac{\Delta\omega_{d_i}}{2}T_E\right)}{\sin\left(\frac{\Delta\omega_{d_i}}{2}T_s\right)}
\label{eq:exp_corr2}
\end{equation}
Our final goal is to track the value of the carrier phase. To this end, the delta-phases obtainable from equation \eqref{eq:exp_corr2} are accomulated as follows:
\begin{equation}
\phi_{i+1}-\phi_i\,\doteq\,\Phi_{i+1}\,=\,Im\left(\log{\frac{C_{i+1}}{C_i}}\right)\,=\,-\Delta\omega_{d_i} T_E.
\label{eq:delta_phi}
\end{equation}
This equation is true if $\Delta\omega_{d_{i+1}}\approx\Delta\omega_{d_i}$, which is a good approximation since the time during which this variation is measured is the coherent time of integration, i.e. in the order of milliseconds.\\

The main advantage of using this \it delta-phase approach \rm to phase tracking is that we have to deal with differential value, that is smaller values than the full phase values. This allows an easy detection of the $\pi$ radiants change of phase due to the presence of the navigation bit.\\

\begin{figure}[tbhp]
\centering
        \includegraphics[width=6cm]{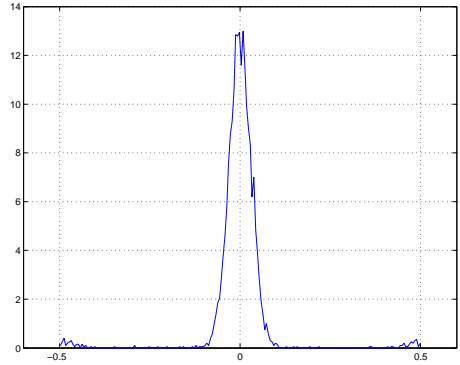} 
    \caption{The histogram of the carrier phase variation, measured on an integration time interval (in this case 1 ms). The x-axis represents cycles, while on the y-axis there are arbitrary units. It is evident the accumulation of $\delta$-phase values around zero, as well as in the vicinity of $\pm$ half a cycle.}
  \label{fig:example_deltaphase_hist}
\end{figure}

\begin{figure}[tbhp]
\centering
        \includegraphics[width=6cm]{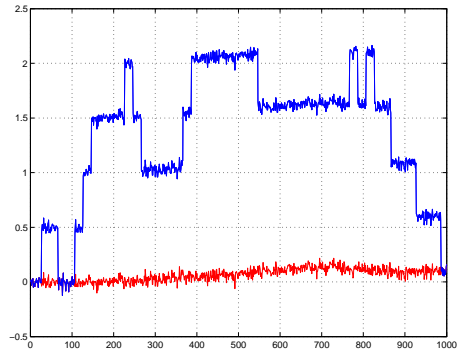} 
    \caption{The carrier phase obtained accumulating the $\delta$-phase according to equation \eqref{eq:delta_phi}. The stepped plot represent this accumulated phase \em as it is \em, i.e. without compensating for the navigation bit half-cycle variation. The lower curve represent the same phase after removal of this effect. The unit of the x-axis is milliseconds, on the y-axis is cycles.}
  \label{fig:example_phase_uphase}
\end{figure}

\begin{figure}
  \centering
    \subfigure[Direct field.]
    {\includegraphics[width=6cm]{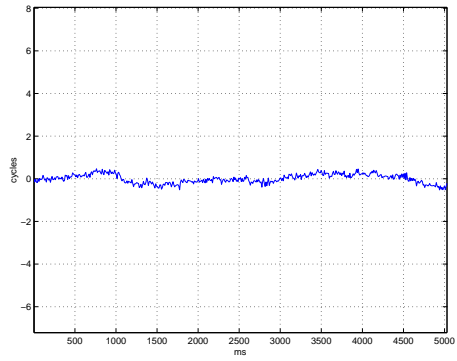}
      \label{fig:casab_phase_detrend_dir}}
    \subfigure[Reflected field.]
    {\includegraphics[width=6cm]{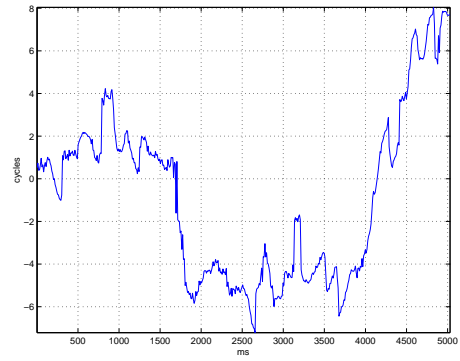}
      \label{fig:casab_phase_detrend_ref}}
  \caption{Example of tracked phase, without the Doppler contribution. The units are milliseconds on the x-axis and cycles on the y axis. The integration time is 20 ms.}
  \label{fig:casab_phase_detrend}
\end{figure}

\begin{figure}
  \centering
    \subfigure[Direct field.]
    {\includegraphics[width=6cm]{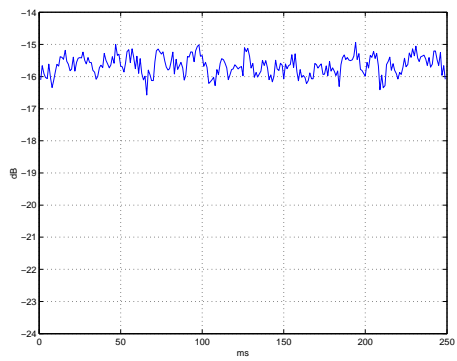}
      \label{fig:casab_ampl_dir}}
    \subfigure[Reflected field.]
    {\includegraphics[width=6cm]{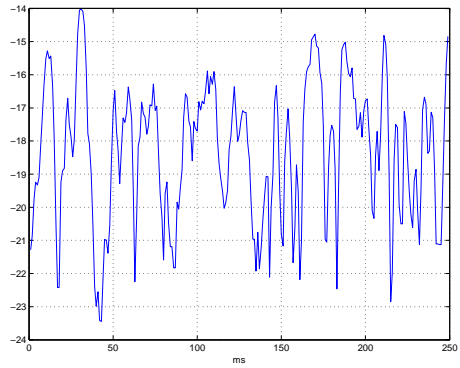}
      \label{fig:casab_ampl_ref}}
  \caption{Example of field amplitude time series. The units are milliseconds on the x-axis and correlation units in dB on the y axis. The integration time is 20 ms.}
  \label{fig:casab_ampl}
\end{figure}

\begin{figure}
  \centering
    \subfigure[Direct field.]
    {\includegraphics[width=6cm]{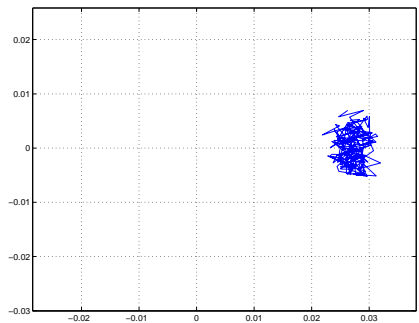}
      \label{fig:casab_ampl_detrend_dir}}
    \subfigure[Reflected field.]
    {\includegraphics[width=6cm]{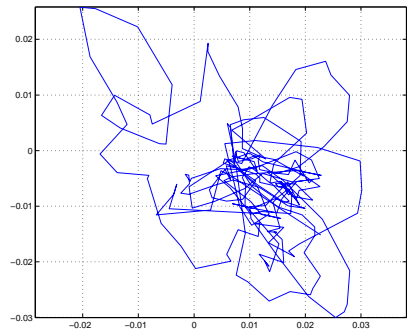}
      \label{fig:casab_ampl_detrend_ref}}
  \caption{Example of complex field time series. The units are correlation units on both axis. The integration time is 20 ms.}
  \label{fig:casab_ampl_detrend}
\end{figure}

\begin{figure}[tbhp]
\centering
        \includegraphics[width=10cm]{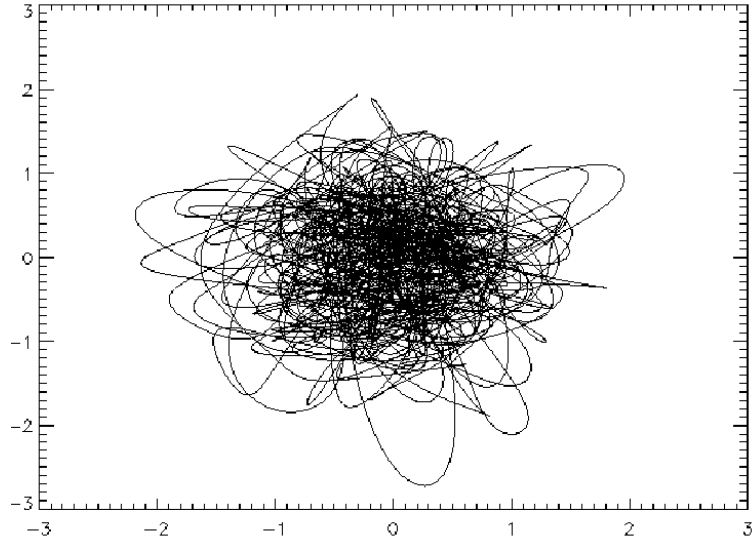} 
    \caption{The EM field at L1 frequency, after reflection on the sea surface. This is a simulated field. The simulation has been performed with a software \cite{ruffini2001b} developed by Starlab. The simulation considers a wind speed $U_{10}=3$ m/s.}
  \label{fig:pacos_dustball}
\end{figure}

Figures from \ref{fig:example_deltaphase_hist} to \eqref{fig:pacos_dustball}, used to explain these concepts, refers to the processing of another set of GPSR data (the Casablanca Experiment). In Figure (\ref{fig:example_deltaphase_hist}) the histogram of the delta-phases is shown. The x-axis represent cycles and the y-axis just arbitrary units. The most part of $\delta$-phase values is clearly concentrated around zero. Another considerable amount of values is allocated just before $\pm\pi$. These values represent in fact \it small \rm values to which $\pm\pi$ radiants have been added, for the occurrence of a navigation bit. 

In Figure (\ref{fig:example_phase_uphase}) both the phase with and without the navigation bit effects are plotted.
In order to derive geophysical parameters from GNSS-R, the phase has to be considered without any Doppler contribution. In Figure \eqref{fig:casab_phase_detrend} this phase for the direct and the reflected signal are shown. The effect of the reflection on the sea surface is clearly visible in the huge variations present in Figure \eqref{fig:casab_phase_detrend_ref} with respect to Figure \eqref{fig:casab_phase_detrend_dir}.
In Figure \eqref{fig:casab_ampl} and \eqref{fig:casab_ampl_detrend} the amplitude magnitude and respectively the complex vector of the direct and reflected fields are shown. In Figure \eqref{fig:pacos_dustball} a simulation of the L1 GPS field after reflection is shown. The parameter of the simulation are chosen according to the Casablanca experiment characteristics.

\section{A {\sc parfait \rm } approach}
\label{ch:diff_approach}
\subsection{Introduction}
It is know that the major inconvenient while tracking the reflected field phase is the occurrence of fadings. Even though in the Bridge 2 experiment, due to the relative smoothness of the sea surface \cite{belmonte2002}, these problematic events are not so frequent, nonetheless, since a single event can completely destroy the phase information, some countermeasures must be taken.
For example (see \cite{belmonte2002}), it is possible to inject in the system (during the fading events) a model-based phase. Another possibility is not to track the reflected field phase at all but to track the direct field minus the reflected field phase, which is really what PARIS concept is about. \\

Let us see more in detail why we should not try to track the phase at all.
First, a distinction should be made. Using the accumulated phase for altimetry presupposes  two difficult tasks:
\begin{itemize}
\item Tracking the phase (hard due to bandwidth and fadings).
\item (If tracking is possible) use the tracked phase for altimetry.
\end{itemize}
Note that the second point is often overlooked. Yet we know that the accumulated phase acts like a random walk. The average accumulated phase is very badly behaved (essentially meaningless).

Tracking the phase is important if the signal of interest rotates fast as compared to noise-induced rotation. This is definitely not the case in the bridge phase or in any other coastal application of the technique. In fact, it is never true for PARIS applications over the oceans. It is also very difficult to do: errors accumulate fast, as above mentioned for example, during fading events.

Even more important, the accumulated phase is meaningless. As we have shown in previous work, the phase wanders around, jumping to different winding number kingdoms. This would appear to be a disaster, but it is not: what we would like to get, if the geophysical signals of interest move slowly enough, is the average field. This exists even if the phase wanders and wanders.

In the bridge experiment the sea state was very smooth. In such a case, tracking the phase may still be useful but in general, this is definitely not true.

\subsection{Using the direct minus reflected combination}
What we are after is  the difference of direct and reflected signals. Note that the same $\omega^E$ will be used for direct and reflected---this simplifies things. We can then return to equation \eqref{eq:delta_phi}, and write
\begin{equation}
\label{eq:interf_field}
 \mbox{Im}\left[\ln C_{i}^R- \ln C_{i}^D)\right] =  d\phi_i^{R-D}=-\Delta\omega_{d_i}^{R-D} T_E.
\end{equation}
Note that we are measuring directly the interferometric field. Let us now address some of important points. 
\subsection{Why we don't need to worry about emission versus reception time differences}
The difference between direct and reflected paths is much less than a microsecond (300 light-meters). Even for a fast moving GPS satellite, this is essentially zero. 
 
This is the essential measurement we are making on reception of a GPS signal:
\begin{equation}
c\Delta t = || x_R(t_R)- x_E(t_E)|| \approx || x_R(t_R)- x_E(t_R) -V_E\delta t  ||,
\end{equation}
where $\delta t $ is the time it take from emission to reception. This is of the order 2x$10^7$ light meters (about 0.07 seconds). Definitely significant: the emitter will move about 300 meters in this time. What does it mean? Well, since we will work with the direct minus reflected travel times, the effect is null. Recall that from a ground platform we essentially have $\Delta_{D_R} \approx 2 h \sin \alpha$, where $\alpha$ is the elevation. In 0.07 seconds, the change of elevation is negligible. \\

On the base of these considerations, a new approach to PARIS altimetry is described in the following sections. It is called \it{PARis Filtered-field AltImetric Tracking} \rm technique (PARFAIT technique) and it proved to be a very robust and precise processing method.\\
\section{The \sc parfait \rm  processing}
\label{parfait_proc}
In this section the basis of the \sc parfait \rm  processing are presented. The \sc parfait \rm  processing is an innovative processing for GNSS-R data developed by Starlab, as part of the STARLIGHT GNSS-R software package. In the frame of this CCN, the application of such a processing will lead to the estimation of the height of the antennas of the Bridge 2 Experiment over the sea surface.\\

Simple geometrical considerations lead to the following equation which relates the height of the receiver over the reflecting surface (considering the same height for the upward looking antenna and for the downward looking antenna) with the delay measured between the direct and reflected GPS signals:
\begin{equation}
delay_{PRN}\left(t\right)\cdot c\,=\,2\,h\left(t\right)\,\sin\left(\epsilon_{PRN}\left(t\right)\right)+d_{offset}
\label{eq:delayVSh}
\end{equation}
where $delay_{PRN}(t)$ is the measured delay between the two signals at time $t$, $h(t)$ is the height of the bridge at time $t$, $\epsilon_{PRN}(t)$ is the elevation at time $t$ of the GPS satellite with a specific PRN number, and $d_{offset}$ is the hardware-induced delay, considered to be constant in time ($c$ is the light speed).
From this equation it is evident that a first estimation of the height of the receiver can be easily done with a linear fit of the measured delay with respect to the sinus of the elevation angles of each satellite.
The delay between direct and reflected signal, though, can be measured through two quantities: C/A code and carrier phase. In the case of carrier phase, equation \eqref{eq:delayVSh} must be rewritten as follows
\begin{equation}
delay_{PRN}\left(t\right)\,=\,2\,h\left(t\right)\,\sin\left(\epsilon_{PRN}\left(t\right)\right)+N_{PRN}\lambda+d_{offset}
\label{eq:delayVSh_phase}
\end{equation}
where this time $delay_{PRN}\left(t\right)$ is already in meters and $\lambda$ is the carrier wavelength. In other words, the equation of each satellite contains an unknown parameter $N_{PRN}$. In order to use all the satellites for one height estimation, it is then necessary to estimate also $N_{PRN}$, i.e. to solve the ambiguity problem. In practise, the following procedure was used. All the measured $delay_{PRN}(t)$ have been put in an unique vector, with an additional additive parameter $b_{PRN}$
\[ (delay_{1}(t_1),\,delay_{1}(t_2),\,...,\:(delay_{2}(t_1)+b_2), ... \]
\[ \, (delay_{2}(t_2)+b_2), \:...,\:(delay_{p}(t_m)+b_p)).\]

 The corresponding vector for the sine of the elevation angles was then built. Using these two vectors, a linear fit has been performed, resulting in a certain value for the norm of the residuals of the fit. Finally, considering this norm as a function of the parameters $b_{PRN}$, the optimum value $\hat{b}_{PRN}$ that minimises the norm has been found. For this  $b_{PRN}$ one obtains the best interpolation, i.e. the best value for a first estimation of the height of the bridge.\\
This estimation of the height is nonetheless quite rough. Let us see why and how to improve this estimation.
First of all, remember that the term on left-hand side of equation \eqref{eq:delayVSh_phase} can be written as (see also equation \eqref{eq:interf_field})
\begin{equation}
delay_{PRN}\left(t\right)\,=\,\lambda d\phi_i^{R-D}
\end{equation}
where $d\phi_i^{R-D}$ is the phase of the (reflected minus direct) interferometric field. This field is generally corrupted by fadings and, during fadings, the phase of this interpherometric signal is completely impossible to track (basically, during these events, the reflected signal is not present at all). Moreover, the optimisation previously described is performed in the $\mathcal{R}$ domain, without correctly approaching the ambiguity problem which is inherently to be solved in the integer domain.\\
To face the fading problem, we propose to filter the interferometric field. This filtering should be long enough to eliminate fading problems and short enough to let the signal of interest pass through. Consider the usual equation
\begin{equation}
\lambda d\phi_i^{R-D}\left(t\right)\,=\,2\,h\left(t\right)\,\sin\left(\epsilon_{PRN}\left(t\right)\right)+d_{offset}.
\end{equation}
The value of $d\phi_i^{R-D}$ should not change for more than a fraction of $2\pi$ in the time duration of the filter. This maximal time depends clearly on the elevation angle of the satellite and, just slightly, form the tide motion. In the case under examination the maximum filtering time is around 10 seconds. In other words, in 10 seconds, at least for one satellite, $d\phi_i^{R-D}$ changes of $\frac{\pi}{2}$ radiants. With this filter length, the fadings are not completely eliminated, even though a realistic estimation of the bridge height (and $d_{offset}$) is now possible. To definitely override the fading problems, a longer integration should be performed, without killing the signal ($d\phi_i^{R-D}$) we wish to measure. To this end, it is possible to counter-rotate the interferometric field, using the first guess of the bridge height as explained in the following.
Consider the interferometric field as
\begin{equation}
E_{interf}\,\propto\,\exp\{d\phi_i^{R-D}\}\,=\,\exp\{2\,h\left(t\right) ...
\end{equation}
\begin{equation}
\sin\left(\epsilon_{PRN}\left(t\right)\right)+d_{offset}\}
\end{equation}
and assume a first guess $\hat{h}_b$ for the bridge height (and $\hat{d}_{offset}$ for the offset).
Than the interferometric field is \it down-converted \rm or \it counter-rotated \em using the bridge height initial guess
\begin{equation}
E_{interf}^{crot}\,\propto\,\exp\{2\,h_b\left(t\right)\,\sin\left(\epsilon_{PRN}\left(t\right)\right)+ ...
\end{equation}
\begin{equation}
d_{offset}\}\cdot\exp{-2\,\hat{h}_b\left(t\right)\,\sin\left(\epsilon_{PRN}\left(t\right)\right)+\hat{d}_{offset}}
\end{equation}
obtaining
\begin{equation}
E_{interf}^{crot}\,\propto\,\exp\{2\,\delta h_b\left(t\right)\,\sin\left(\epsilon_{PRN}\left(t\right)\right)+\delta d_{offset}\}.
\label{eq:interf_field_crot}
\end{equation}
Clearly, the phase of the field in equation \eqref{eq:interf_field_crot} varies much slower than the phase of the original interferometric field and this allows a longer filtering time, that is, virtually,  a complete elimination of fadings problems.\\
This technique provides a fundamental cornerstone for PARIS processing from air and spaceborne applications.\\

The equation which relates the phase delay between direct and reflected signal, the satellite elevations and the $\delta h_b$ (i.e. the error between the first guess of the bridge height and the real value) turns out to be
\begin{equation}
\lambda d\phi_i^{R-D}\left(t\right)\,=\,2\,\delta h_b\left(t\right)\,\sin\left(\epsilon_{PRN}\left(t\right)\right)+N_{PRN}\lambda+\delta d_{offset}
\label{eq:delayVSDeltah_phase}
\end{equation}
This is the new equation to be used to fit the straight line and infer the height of the bridge, with respect to the first guess used to counter-rotate the field.
As already said, the equation relating the interferometric phase and the bridge height is known up to a multiple of wavelength (equation \eqref{eq:delayVSh_phase}) or, in other words, there is an ambiguity due to the periodic behaviour of the phase. In order to correctly solve this ambiguity problem, a search is performed in the space of the n-uples of integers and the one that produce the linear fit with smallest residue is assumed as the true one. It is important to point out that the space of the n-uples to be spanned in order to remove the ambiguity is drastically reduced by the filtering of the field previously described. In fact, since the only residual \it movement \rm in the filtered data is the tide, the single \it filtered \rm phase history, will all lie in a narrow height interval. This interval is determined basically by the correctness of first guess for the bridge height. In other words, if the guess is within $\pm$ half meter, the n-uples subspace to be scan can be limited to those n-uples whose components belong to the interval $[-3,3]$, centred on the first guess of the n-uple, obtained from the real solution.\\
Another possibility to reduce the cardinality of the subspace of the n-uples to check, is to consider that satellites with similar elevation angles cannot have very different integer ambiguities.\\

\section{First altimetric results}
\label{ch:first_altimetric_results}
The \sc parfait \rm  analysis described in the previous section, has been applied to the first 10 minutes of the Bridge2 data, part A1 and to the first 10 minutes of the part A2. The following step have been performed accordingly:
\begin{itemize}
\item the EM fields, direct and reflected, have been computed through the GPS correlation process;
\item the   field has been counter-rotated (equation \eqref{eq:interf_field_crot});
\item the   field has been filtered;
\item the phase of the  field has been evaluated;
\item a straight line has been interpolated to the phase histories (one for each visible satellite) against the elevation angle (equation \eqref{eq:delayVSDeltah_phase});
\item the real n-uple of values of $\lambda N_{PRN}$ that minimise the residue of the fit is use to define a first guess for the integer n-uple of values of $N_{PRN}$;
\item the linear fit has been again performed for every integer n-uple of $N_{PRN}$ to find the one that minimises the residuals.
\end{itemize}

\begin{figure}[tbhp]
\centering
 \includegraphics[width=10cm]{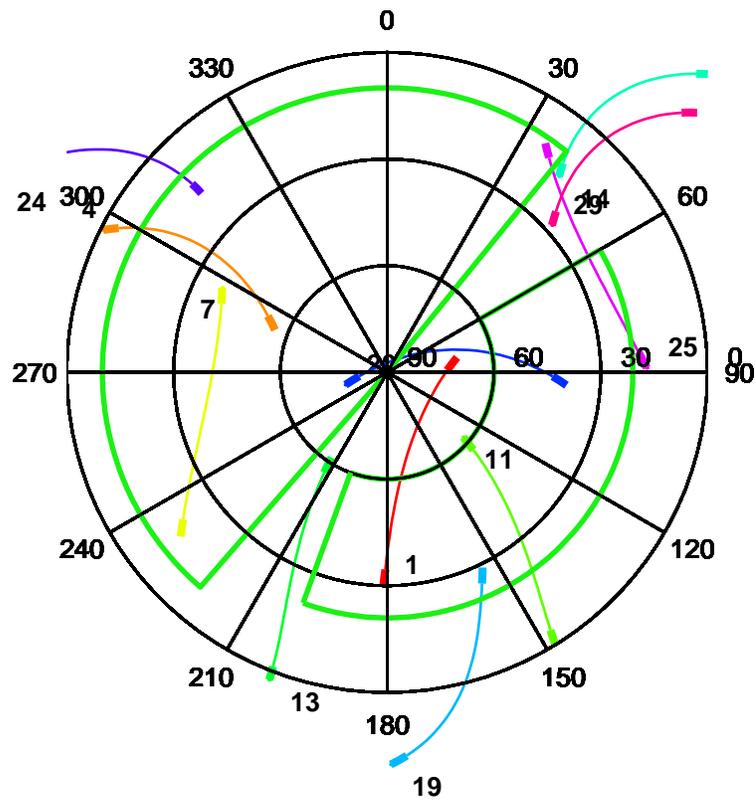}
 \caption{Each coloured arc represents the position of a GPS satellite from the start of the part A1 of the experiment to the beginning of part A2 plus 10 minutes. The PRN number is closed to the beginning of the data. The green mask represent the area where the GPS signal reflections are supposed to be free of shadowing phenomena due to the bridge structure, and therefore only the satellite within this mask can be taken into consideration for PARIS processing. The bold parts of the lines represent the first and the second 10 minutes periods.}
\label{fig:zeeland_sat}
\end{figure}

\begin{figure}
  \centering
    \subfigure[In this figure, the reflected-minus-direct phase delay for each PRN is  plot with $N_{PRN}=0$.]
    {\includegraphics[width=8cm]{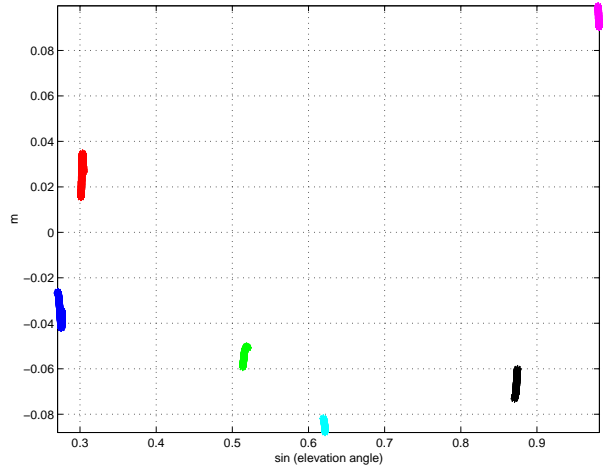}
      \label{fig:phases_hist_example_1}}
    \subfigure[In this figure, the reflected-minus-direct phase delay for each PRN is  plot with $N=\{0 0 1 1 2 3\}$.]
    {\includegraphics[width=8cm]{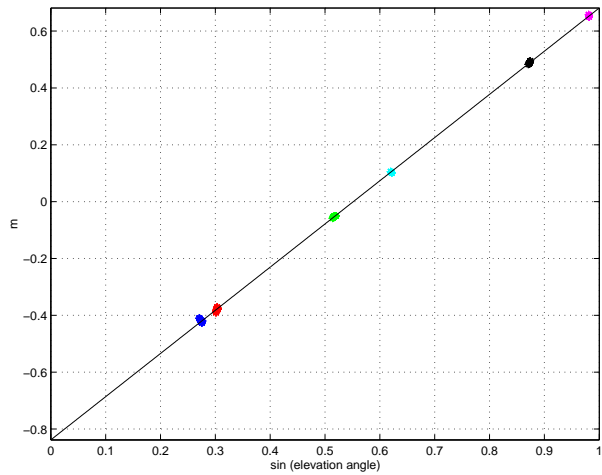}
      \label{fig:phases_hist_example_2}}
  \caption{Each coloured spot represents the reflected-minus-direct phase delay VS satellite-elevation for a certain PRN number.}
  \label{fig:SNR_examples}
\end{figure}

The filter used for smoothing the field is a flat zero-phase filter, with a length of 30 seconds.\\
This filtering has been performed for almost\footnote{Satellites outside the \it Zeeland Mask \em, as defined in \cite{belmonte2001} are discharged (see also caption of Figure \eqref{fig:zeeland_sat}.} all visible satellites (see table (\ref{tab:visible_sat1}) and Figure \eqref{fig:zeeland_sat}). The phase histories are shown in Figure \eqref{fig:phases_hist_example_1}. A straight line has been fit through these phase histories, against the sinus of the satellite elevation angle (figure \eqref{fig:phases_hist_example_2}).

This fitted line gives an estimation of the bridge height of 18.61 m, an hardware bias of -0.81 and, as first guess for the n-uple that solves the ambiguity problem, the values $[0\:0\: 1 \:1\: 2\: 3]$. Now, a search in a subset of $\mathcal{I}^6$ is done to minimise the residuals of the fit in the space of the n-uples of integers. The subspace considered is the one spanned by all the combination of integers between $\pm3$ around the first guess. The result is the n-uple $[0\: 0\: 2 \:2 \:4\: 5]$ which gives a bridge height estimation of $18.82$ m and an instrumental bias of $-0.45$. 

\begin{table}
\vspace{.5cm}
\label{tab:visible_sat1}
\centering{
\begin{tabular}{|r|c|c|c|}
  \hline
 PRN & elevation & \parbox{2cm}{\centering{mean $SNR$ (direct)}} & \parbox{2cm}{\centering{mean $SNR$ (reflected)}} \\
        \hline \hline
14  & 17$^o$   & 29.4 dB& 25.0 dB\\
  \hline                  
25  & 17$^o$   & 32.0 dB& 25.8 dB \\
  \hline                  
1  &  30$^o$   & 31.2 dB& 24.6 dB\\
  \hline                  
7  &  38$^o$   & 33.2 dB& 29.4 dB\\
  \hline                  
11  & 62$^o$   & 34.0 dB& 29.4 dB\\
  \hline                  
20  & 78$^o$   & 30.4 dB& 26.6 dB\\
      \hline
\end{tabular}}
\caption{Visible satellites, their elevation in degrees, the 10 ms coherent integration mean $SNR_{dB_w}$ for the direct and the reflected signal.}
\end{table}

\begin{figure}
  \centering
    \subfigure[The upper line is the bridge height estimated with the procedure explained in this chapter. The bottom line is the bridge height according to the available tide measurements and the GMV calibration of the absolute height at 14:57:27 UTC, using GPS code differential processing \cite{belmonte2002}.]
    {\includegraphics[width=8cm]{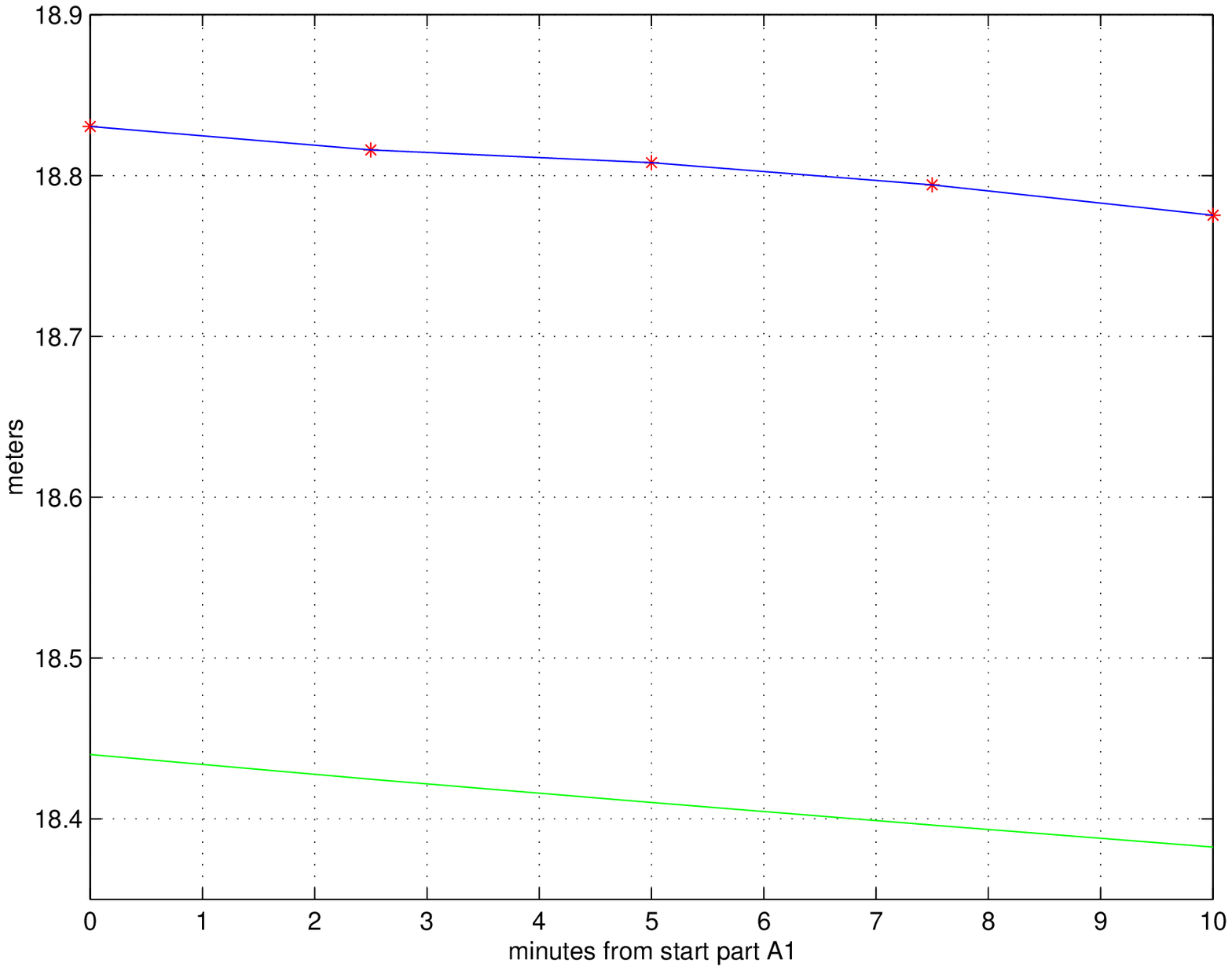}
      \label{fig:bridge_cfr_partA1_1}}
    \subfigure[In this plot, the estimations of the height (red *) and the actual height (line) are shown, after adding to the actual height the mean of the values of the last column of Table \eqref{tab:height_results_1}.]
    {\includegraphics[width=8cm]{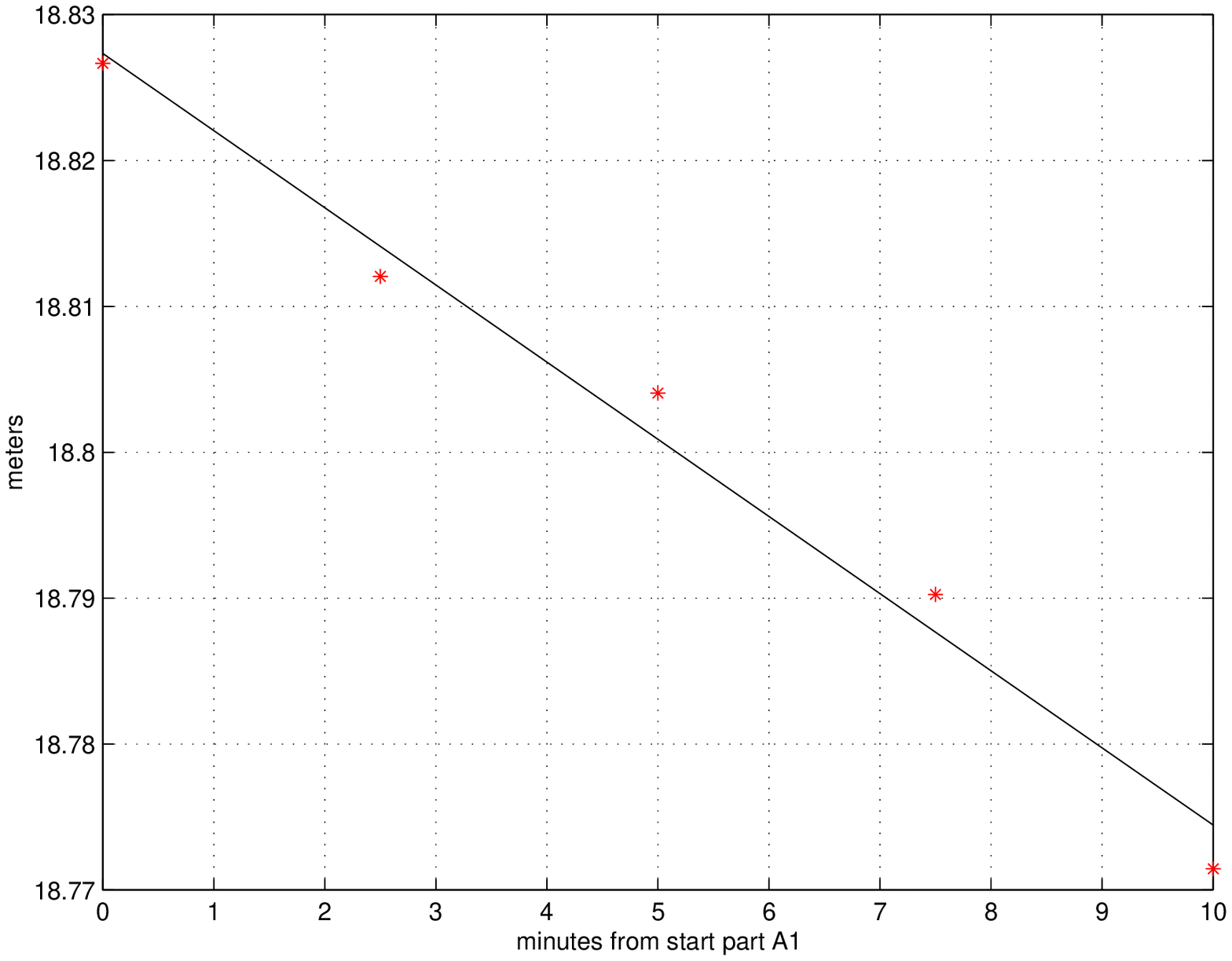}
      \label{fig:bridge_cfr_partA1_2}}
  \caption{The bridge height estimation during the first 10 minutes of the part A1 of the data.}
  \label{fig:bridge_cfr_partA1}
\end{figure}

\begin{figure}
  \centering
    \subfigure[The upper line is the bridge height estimated with the procedure explained in this chapter. The bottom line is the bridge height according to the available tide measurements and the GMV calibration of the absolute height at 14:57:27 UTC, using GPS code differential processing \cite{belmonte2002}.]
    {\includegraphics[width=8cm]{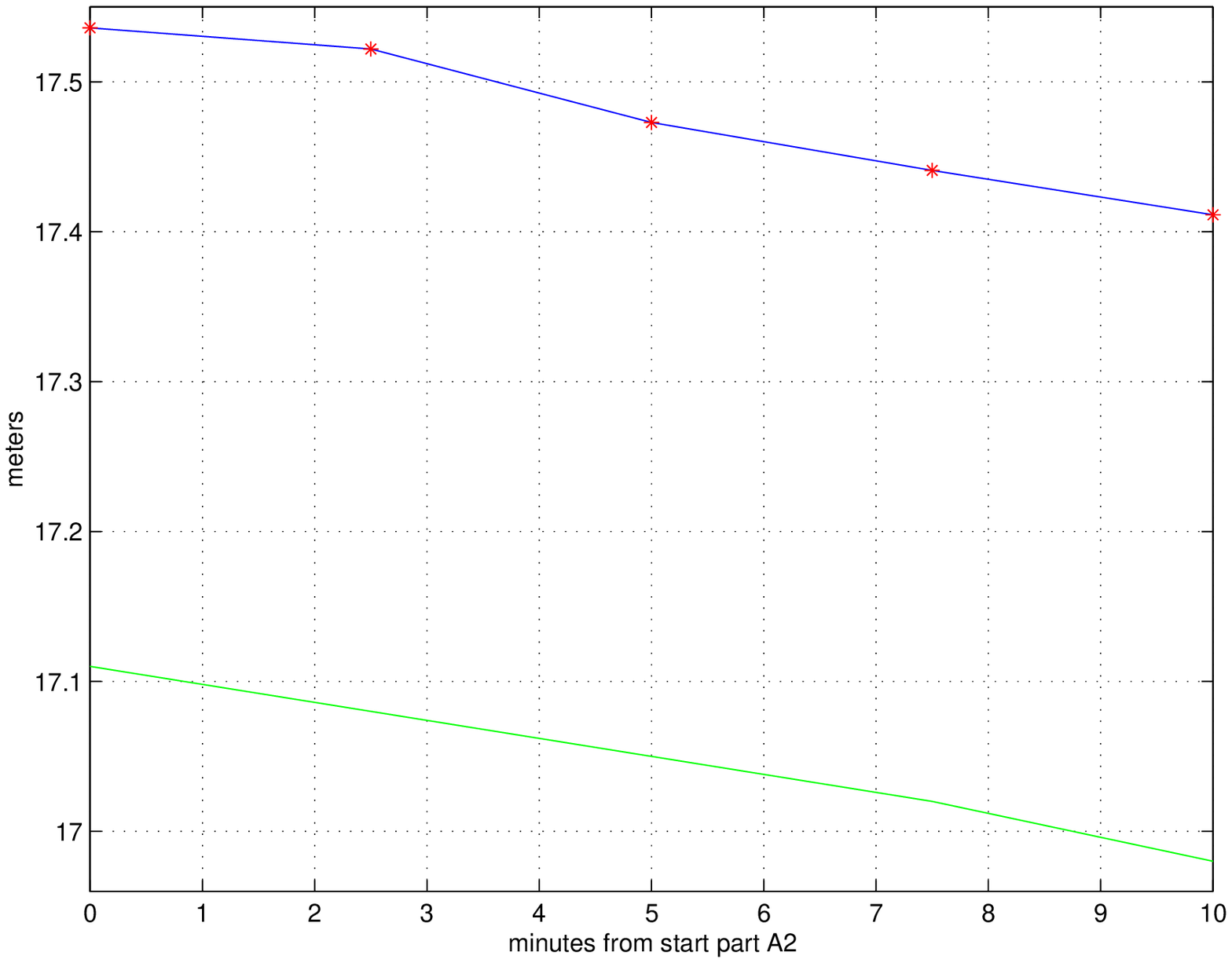}
      \label{fig:bridge_cfr_partA2_1}}
    \subfigure[In this plot, the estimations of the height (red *) and the actual height (line) are shown, after adding to the actual height the mean of the values of the last column of Table \eqref{tab:height_results_2}.]
    {\includegraphics[width=8cm]{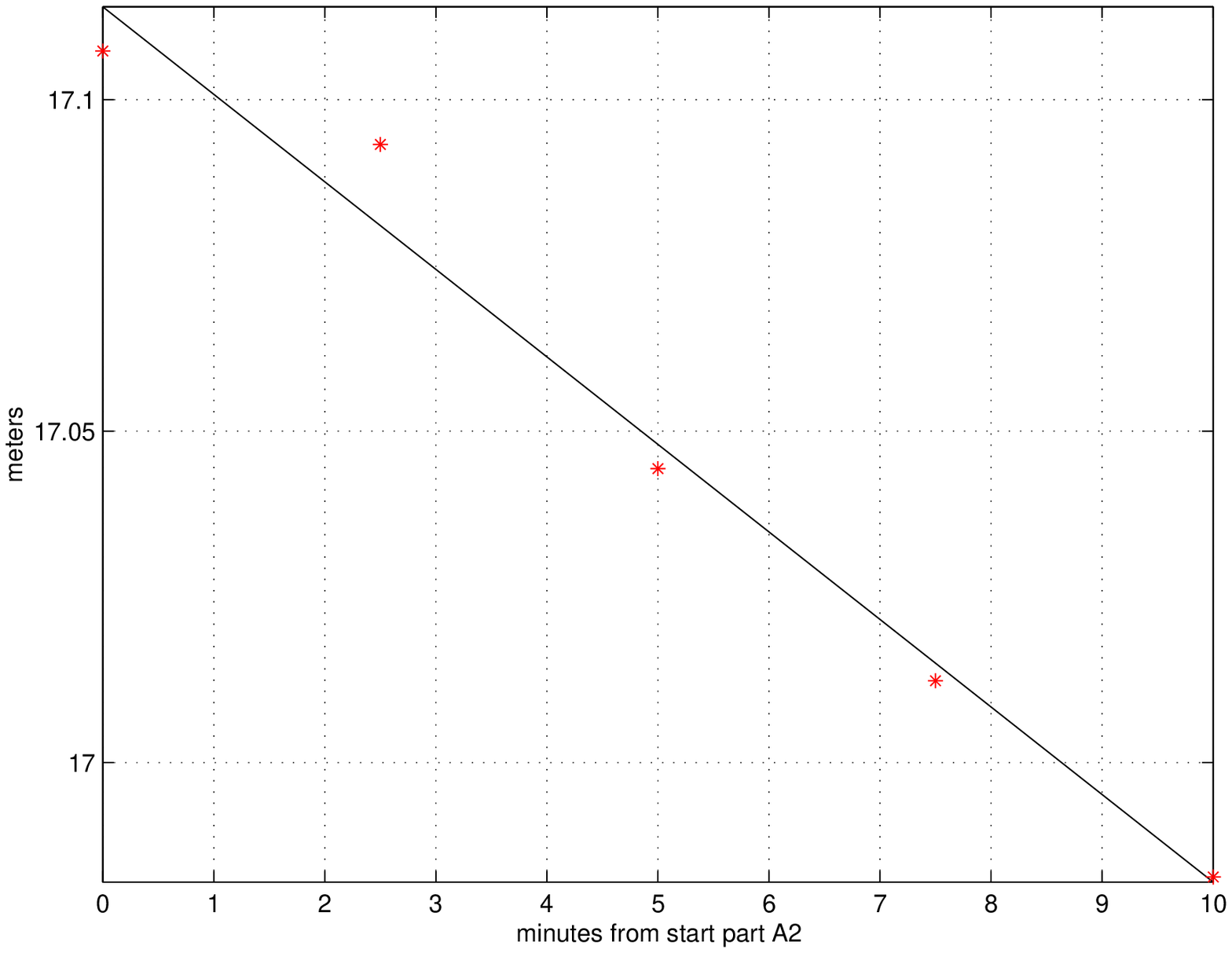}
      \label{fig:bridge_cfr_partA2_2}}
  \caption{The bridge height estimation during the first 10 minutes of the part A2 of the data.}
  \label{fig:bridge_cfr_partA2}
\end{figure}

\begin{figure}[t!]
 \centering
\label{fig:true_bh_vs_est_bh}
 \includegraphics[width=8cm]{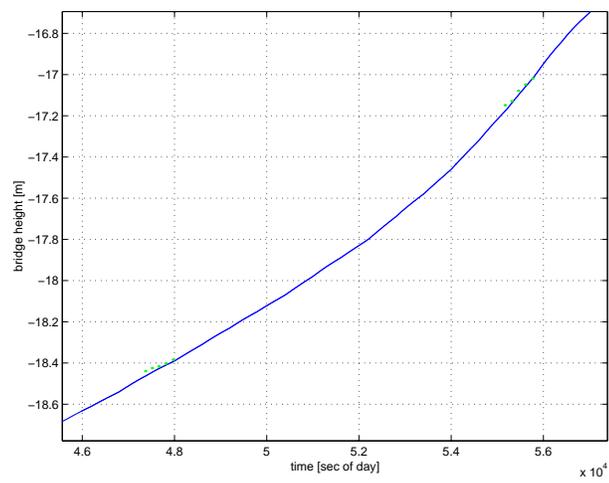}
 \caption{ The solid line is the distance between the up-looking antenna and the sea surface, according to the available tide measurements and the GMV calibration of the absolute height at 14:57:27 UTC, using GPS code differential processing \cite{belmonte2002}. The green dots are the estimated values, after removing the bias.}
\end{figure}

This procedure has been applied to the first 10 minutes of the part A1 and of part A2 of the data. The results are reported in Table \eqref{tab:height_results_1} and in Figure \eqref{fig:bridge_cfr_partA1} for part A1 and in Table \eqref{tab:height_results_2} and in Figure \eqref{fig:bridge_cfr_partA2} for part A2. The final value of the estimated bridge height is considered to be the interpolated straight line through the available point \it after removing the supposed estimation bias \rm . 
The use of a straight line for the interpolation is justified by the short period of time considered, relatively to the tide period. The standard deviation of the interpolated estimation w.r.t. the measured tide is of $0.35$ cm for part A1 and of $0.84$ cm for part A2. The results for both periods are plotted in figure \eqref{fig:true_bh_vs_est_bh}.\\

Fitting both parts to the tide curve, i.e. choosing the bias that minimises the standard deviation of the data to tide ``ground truth'', leads to a bias of 41.10 cm and a standard deviation of 1.81 cm. This bias could be due either to an error in the determination of the absolute value of the height of the bridge performed with the differential GPS processing or, partially, to some anomalies in the flowing of the water in the vicinity of the bridge structures.\\

Moreover, assuming also that the tide measured below the bridge can have a time delay with respect to the place were the tide is measured, the best fit (over both bias and delay) is obtained with a delay of 3 minutes and 12 seconds with respect to the time of the tide data collection and with a bias of 39.13 cm. The standard deviation of the fitted data with respect to the tide curve is in this case of 0.893 cm. \\

\begin{table}[ht]
\label{tab:height_results_1}
\centering{
\begin{tabular}{|c||c||c|c||c|}
  \hline
 \parbox{2cm}{\centering{time (minutes from start)}} & \parbox{2cm}{\centering{instrumental bias [cm]}} &\parbox{2.5cm}{\centering{bridge height estimation [m]}} &\parbox{2cm}{\centering{assumed ground truth [m]}} & \parbox{2cm}{\centering{difference [cm]}} \\
\hline \hline
1      & -0.45 &  18.83   &  18.44  & 39.06 \\
\hline                  
2      & -0.45 &  18.82   &  18.42  & 39.13 \\
\hline                  
3      & -0.46 &  18.81   &  18.41  & 39.78 \\
\hline                  
4      & -0.45 &  18.79   &  18.40  & 39.80 \\
\hline                  
5      & -0.26 &  18.78   &  18.38  & 39.29 \\
\hline 
\end{tabular}}
\vspace{.5cm}
\caption{Results of the bridge height estimation during the first 10 minutes of the part A1 data. }
\end{table}

\begin{figure}[t!]
 \centering
\label{fig:crf_with_tide_2}
 \includegraphics[width=8cm]{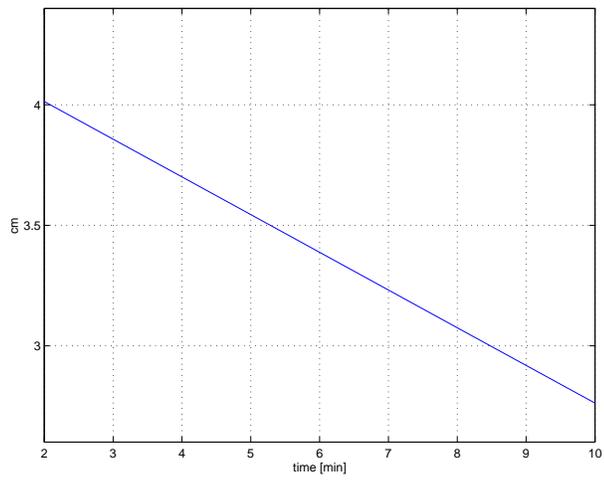}
 \caption{This line represents the difference of two differences: the difference of the measured bridge heights during the collection of the first part of data and those measured during the second part minus the same difference but calculated with the estimated data. The result confirms again the effectiveness of the \sc parfait \rm  processing.}
\end{figure}

Another interesting check to be done to asses the validity of the processing is to compare the change in the height of the bridge between the first 10 minutes of part A1 and the first 10 minutes of part A2 as retrieved by the processing and as provided by the tide measurements. This comparison is shown in figure \eqref{fig:crf_with_tide_2}. The result is absolutely satisfying: the estimation is in accordance with the measures within about 4 cm.\\

 To summarise, the proposed approach to PARIS altimetry, the \sc parfait \rm  technique, leads to possibly biased but very precise estimation of the tide, 
\begin{itemize}
\item without the need to insert any kind of model for the phase of the reflected signal during fadings;
\item without rejecting too many visible satellites because of their poor SNR and/or frequent fadings;
\end{itemize}
\begin{table}[ht]
\label{tab:height_results_2}
\centering{
\begin{tabular}{|c||c||c|c||c|}
  \hline
 \parbox{2cm}{\centering{time (minutes from start)}} & \parbox{2cm}{\centering{instrumental bias [cm]}} &\parbox{2.5cm}{\centering{bridge height estimation [m]}} &\parbox{2cm}{\centering{assumed ground truth [m]}} & \parbox{2cm}{\centering{difference [cm]}} \\
\hline \hline
1      & -0.27 & 17.54    & 17.11   & 42.6 \\
\hline         
2      & -0.28 & 17.52    & 17.08   & 44.2 \\
\hline         
3      & -0.26 & 17.47    & 17.05   & 42.3 \\
\hline         
4      & -0.08 & 17.44    & 17.02   & 42.1 \\
\hline         
5      & -0.08 & 17.41    & 16.98   & 43.1 \\
\hline 
\end{tabular}}
\vspace{.5cm}
\caption{Results of the bridge height estimation during the first 10 minutes of the part A2 data. }
\end{table}


\section*{Acknowledgements} 

The authors wish to thank Manuel Martin-Neira (ESA-ESTEC) and Maria Belmonte (ESA-ESTEC) for useful discussions and real (and on-going) collaboration.


\bibliographystyle{plain}
\addcontentsline{toc}{chapter}{Bibliography}
\bibliography{/home/alkaid/intranet/library/feosbiblio}

\begin{thebibliography}{10}

\bibitem{armatys2000}
M.~Armatys, A.~Komjathy, P.~Axelrad, and S.~Katzberg.
\newblock A comparion of {GPS} and scatterometr sensing of ocean wind speed and
  direction.
\newblock In {\em Proc. IEEE IGARSS, Honolulu, HA}, 2000.

\bibitem{caparrini1998}
M.~Caparrini.
\newblock Using reflected {GNSS} signals to estimate surface features over wide
  ocean areas.
\newblock Technical Report EWP 2003, ESA report, December 1998.

\bibitem{caparrini2001}
M.~Caparrini and G.Ruffini.
\newblock Casablanca data processing.
\newblock Starlab "Knowledge Nugget" kn-0111-001, 2001.

\bibitem{cardellach2001a}
E.~Cardellach, J.M. Aparicio, A.~Rius, J.S., and J.~Torrobella.
\newblock Application of the paris concept to transoceanic aircraft remote
  sensing.
\newblock Technical report, IEEC, 2001.

\bibitem{cardellach2002c}
E.~Cardellach and A.~Rius.
\newblock Comparison of {PAFEX} estimates with ground truth.
\newblock Technical report, IEEC, 2002.
\newblock WP210 of OPPSCAT 2 Project - ESA contract RFQ/3-10120/01/NL/SF.

\bibitem{cardellach2002b}
E.~Cardellach and A.~Rius.
\newblock Inversion of {PAFEX} data with elfouhaily's technique.
\newblock Technical report, IEEC, 2002.
\newblock WP205 of OPPSCAT 2 Project - ESA contract RFQ/3-10120/01/NL/SF.

\bibitem{cardellach2002}
E.~Cardellach and A.~Rius.
\newblock Preprocessing of {PAFEX} data.
\newblock Technical report, IEEC, 2002.
\newblock WP120 of OPPSCAT 2 Project - ESA contract RFQ/3-10120/01/NL/SF.

\bibitem{cardellach2001}
E.~Cardellach, G.~Ruffini, D.~Pino, A.~Rius, A.~Komjathy, and J.~Garrison.
\newblock Mediterranean balloon experiment: Gps reflection for wind speed
  retrieval from the stratosphere.
\newblock {\em submitted to Remote Sensing of Environment}, 2001.

\bibitem{garrison1996}
J.L. Garrison, S.J. Katzberg, and C.T. Howell.
\newblock Detection of ocean reflected gps signals: theory and experiment.
\newblock In {\em IEEE Southeaston '97}. IEEE, April 1997.

\bibitem{garrison2000}
J.L. Garrison, G.~Ruffini, A.~Rius, E.~Cardellach, D.~Masters, M.~Armatys, and
  V.U. Zavorotny.
\newblock Preliminary results from the gpsr mediterranean balloon experiment
  (gpsr-mebex).
\newblock In {\em Proceedings of ERIM 2000}, Charleston, South Carolina, USA,
  May 2000.

\bibitem{garrison1998}
L.~Garrison, S.~Katzberg, and M.~Hill.
\newblock Effect of sea roughness on bistatically scattered range coded signals
  from the {GPS}.
\newblock {\em Geophysical Research Letters}, 25:2257--2260, 1998.

\bibitem{garrison2000a}
L.~Garrison, S.~Katzberg, V.~Zavorotny, and D.~Masters.
\newblock Comparison of sea surface wind speed estimates from reflected {GPS}
  signals with buoy measurements.
\newblock In {\em Proc. IEEE IGARSS, Honolulu, HA}, 2000.

\bibitem{germain2002}
O.~Germain and G.~Ruffini.
\newblock Least square inversion of {PAFEX} data.
\newblock Technical report, Starlab Barcelona SL, 2002.
\newblock WP200 of OPPSCAT 2 Project - ESA contract RFQ/3-10120/01/NL/SF.

\bibitem{pipaer2000a}
G.Ruffini and F.Soulat.
\newblock Paris interferometric processor theoretical feasibility study part i
  and part ii.
\newblock Technical report, ESA contract 14071/99/nl/mm, 2000.

\bibitem{komjathy1998}
A.~Komjathy.
\newblock Gps surface reflection using aircraft data: analysis and results.
\newblock In {\em Proceedings of the GPS surface reflection workshop}. Goddard
  Space Flight Center, July 1998.

\bibitem{labrecque1998}
J.~LaBrecque, S.T. Lowe, L.E. Young, E.R. Caro, L.J. Romans, and S.C. Wu.
\newblock The first spaceborne observation of gps signals reflected from the
  ocean surface.
\newblock In {\em Proceedings IDS workshop}. JPL, December 1998.

\bibitem{martin-neira2001}
M.~Mart\'{i}n-Neira, M.~Caparrini, J.~Font-Rossello, S.~Lannelongue, and
  C.~Serra.
\newblock The paris concept: An experimental demonstration of sea surface
  altimetry using gps reflected signals.
\newblock {\em IEEE Transactions on Geoscience and Remote Sensing},
  39:142--150, 2001.

\bibitem{pipaer2000}
PIPAER.
\newblock Paris interferometric processor analysis and experiment results.
\newblock Technical report, IEEC and GMV - ESA contract 14071/99/nl/mm, 2000.

\bibitem{belmonte2001}
M.~Belmonte Rivas and M.~Mart\'{i}n-Neira.
\newblock {GNSS} reflections: First altimetry products from bridge-2 field
  campaign.
\newblock unpublished.

\bibitem{belmonte2002}
M.~Belmonte Rivas and M.~Martin-Neira.
\newblock {GNSS} reflections:first altimetry products from bridge-2 field
  campaign.
\newblock In {\em Proceedings of {NAVITEc}, 1st {ESA} Workshop on Satellite
  Navigation User Equipment Technology}, pages 465--479. ESA, 2001.

\bibitem{ruffini2001b}
G.~Ruffini, M.~Caparrini, and B.~Chapron.
\newblock Improved ocean and em models for in-silico spaceborne {GNSS-R}.
\newblock Technical report, PARIS Beta WP3200 - ESA ESTEC CONTRACT No.
  15083/01/NL/MM, 2001.

\bibitem{ruffini2001a}
G.~Ruffini, M.~Caparrini, O.~Germain, F.~Soulat, and J.~Lutsko.
\newblock Remote sensing of the ocean by bistatic radar observations: a review.
\newblock Technical report, PARIS Beta WP1000 - ESA ESTEC CONTRACT No.
  15083/01/NL/MM, 2001.

\bibitem{ruffinipipaer2000}
G.~Ruffini and F.~Soulat.
\newblock Paris interferometric processor analysis and experiment resultsi,
  http://arxiv.org/physics/0011027.
\newblock Technical report, IEEC and GMV - ESA contract 14071/99/nl/mm, 2000.

\bibitem{zavorotny2000}
V.~Zavorotny and A.~Voronovich.
\newblock Scattering of {GPS} signals from the ocean with wind remote sensing
  application.
\newblock {\em IEEE Trans. Geoscience and Remote Sensing}, 38(2):951--964,
  2000.

\bibitem{zuffada2000}
C.~Zuffada, R.~Treuhaft, S.~Lowe, G.~Haij, M.~Lough, L.~Young, Wu~S, and
  M.~Smith.
\newblock Altimetry with reflected {GPS} signals: results from a lakeside
  experiment.
\newblock In {\em Proceedings IGARSS 2000}, 2000.

\end{thebibliography}

\begin{figure}[tbhp]
\centering
      \scalebox{.5}{
        \includePSTeX{downconvertion} 
        }
    \caption{Starting from the in-phase component of the sampled signal, with a carrier frequency equal to the sum of the IF of the receiver and the Doppler frequency, two downconvertions are performed, with two phasor relatively delayed of $\pi/2$. After a low-pass filtering, the two obtained signals bring information about the amplitude of the backscattered EM field for both the in-phase and in-quadrature components.}
  \label{fig:downconvertion}
\end{figure}


\begin{figure}[t!]
\centering
 \includegraphics[width=8cm]{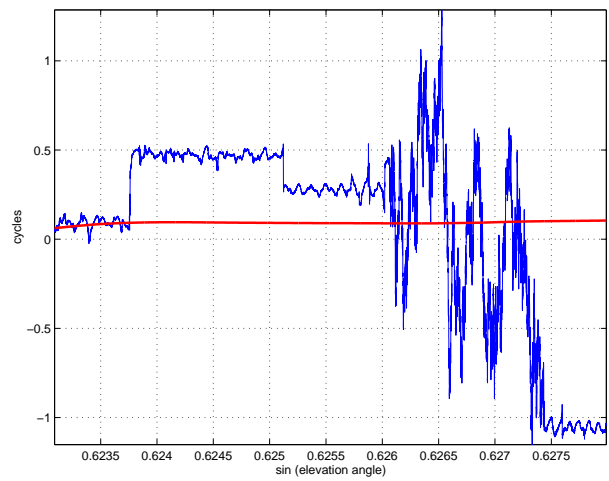}
 \caption{ \label{fig:filtered_phase} The blue curve is the phase of the interferometric field for PRN number 7, from minute 2 to minute 4 of the part A1 of the Bridge 2 experiment. It is clearly visible the occurrence of a long fading (on the left-hand side of the plot) and of isolated phase slips. These phenomena disappear in the phase of the filtered interferometric field (red line).}
\end{figure}


\end{document}
\end